\documentclass{article}

\usepackage{arxiv}

\usepackage[utf8]{inputenc} 
\usepackage[T1]{fontenc}    
\usepackage{hyperref}       
\usepackage{url}            
\usepackage{booktabs}       
\usepackage{amsfonts}       
\usepackage{nicefrac}       
\usepackage{microtype}      
\usepackage{lipsum}

\usepackage{graphicx} 

\usepackage{float}

\usepackage{bm}

\usepackage{epstopdf}

\title{Non-Trivial Topological Phase in the Sn$_{1-x}$In$_x$Te Superconductor}

\author{Tome M. Schmidt \\
Instituto de F\'{\i}sica, Universidade Federal de Uberl\^{a}ndia, \\
Uberl\^{a}ndia, Minas Gerais 38400-902, Brazil
\And
G. P. Srivastava \\
School of Physics, University of Exeter, \\
Stocker Road, Exeter EX4 4QL, UK}

\begin{document}
\maketitle

\begin{abstract}

Whereas SnTe is a inverted band gap topological crystalline insulator,
the topological phase of the alloy Sn$_{1-x}$In$_x$Te, a topological 
superconductor candidate, has not been clearly studied so far.
Our calculations show that the Sn$_{1-x}$In$_x$Te band gap reduces by increasing the In
content, becoming a metal for $x>0.1$.
However, the band inversion at the fcc L point for both gapped  and gapless phases
has been maintained. Furthermore,
the computed topological invariant shows a non-trivial phase 
with a mirror Chern number $n_M  = -2$ for In concentrations of
$x=0.03125$, $x=0.125$, and $x=0.25$. We also identify
pairs  of topologically
protected  states  on  the (001) surface  of Sn$_{1-x}$In$_x$Te  with  $\pm i$
mirror  eigenvalues.  The character of these topological states is affected by 
In dopant.  As  the In  content $x$
increases, the Dirac crossing point  moves further away from the
L point, and the Fermi  velocity of the topological states increases
significantly. Our results demonstrate a non-trivial topological phase for the
superconductor Sn$_{1-x}$In$_x$Te,
and provide a detailed description of the topological state properties.

\end{abstract}


\section{Introduction}

Topological phase of matter has received significant attention lately,
starting from  topological   insulators    (TIs)
and  more recently  topological superconductors  (TSCs)
\cite{RMP82-3045,Ando-JPSJ82-102001,RMP83-1057,ReviewAndo-Fu-2016,ReviewTCI-Ando}.
TIs  are
characterized by Dirac-like edge or surface states that can be protected by
time  reversal symmetry \cite{Kane2005},  or crystal lattice symmetry, the so
called topological crystalline insulator (TCI)
\cite{Fu2011}.   TSCs  has  special interest  for  potential  quantum
computing applications \cite{ReviewTCI-Ando}. Topological superconductor phase
has been proposed to be obtained throw proximity  induced 
by attaching  a TI  onto a
conventional  superconductor   \cite{PRL100-096407,Nat10-943},  or  by
doping a TI turning it into a  superconductor phase, like Cu, Sr or Nb
doped Bi$_2$Se$_3$
\cite{PRL104-057001,PRL106-127004,PRB84-054513,PRB83-220513,JACS137-10512,
  PRB92-020506,PRB91-144506,PRB94-180510,ChemMat28-779}.
Recently, superconductivity  has also been investigated  by the doping
of    a    TCI    material,    in    particular    In    doped    SnTe
\cite{PRL109-217004,PRB79-024520,PRL110-206804,PRB87-140507,EPL108-37010,
  PRB90-064508,NL15-3827,PRB93-024520,PRB96-104502,PRB97-024511,PRB98-054503}
or   PbSnTe  \cite{PRB97-220502}.    As   SnTe  is   a  TCI   material
\cite{HsiehNC-2012,Tanaka2012} and In doped  SnTe is a superconductor,
Sn$_{1-x}$In$_x$Te is expected  to be  a TSC  material.  Indeed,
Sn$_{1-x}$In$_x$Te  is  a  well  characterized  superconductor,  whose
critical  temperature depends  on  the In  doping  rate, varying  from
T$_{\rm{C}}=1.2~K$  in low  In doping  regime \cite{PRL110-206804}  to
T$_{\rm{C}}=4.8~K$   for  higher   In  contents   \cite{PRB90-064508}.
Although  there is  no doubt  about the  superconductivity phase,  the
preservation  of  its  topological   phase  under  In  doping    
is a topic of current discussion.
For low In concentration some experiments indicate
a       signature       of      topological       surface       states
\cite{PRL110-206804,NL15-3827},  and  recent  experiments 
\cite{PRB93-075132} show evidence  of surface linear dispersion
bands  at  high   In  doping  rates.   However,  that does not confirm
its topological order.  

Also the superconductivity mechanism in Sn$_{1-x}$In$_x$Te is currently
in discussion. Some  previous experiments indicate
that  the superconductivity should  be
odd-parity   pairing   \cite{PRL109-217004,PRB87-140507},   but   more
recently   it   has   been   attributed  to   a   s-wave   BCS   model
\cite{EPL108-37010,PRB90-064508,PRB96-104502,PRB97-024511}.    Another
intriguing question  with respect to In  doped SnTe TCI is  its doping
character. While low In  concentrations generate hole-like doping, for
high    In    doses    it   turns    into    electron-like    carriers
\cite{PRB93-024520,PRB98-054503}.  The  hole-like can be  expected, as
In is a group III element substituted at a group IV atom, but no
electron-like behaviour would be expected.

In this  work we  show that
Sn$_{1-x}$In$_x$Te for In concentrations
of  $x=0.03125$, $x=0.125$  and  $x=0.25$ the  valence and  conduction
bands are  inverted, giving a non-null Chern number.
We have identified pairs of topological states protected by the mirror
symmetry  on the (001) surface of  Sn$_{1-x}$In$_x$Te with  opposite mirror
eigenvalues $\pm i$.  Our  results show that the  experimentally observed hole
to  electron carrier-like  transition as  a function  of increased  In
concentration can  be explained  in terms of  the depopulation  of the
In-5s orbital,  which is pinned at  the Fermi level. Furthermore, our  results also
indicate that Sn$_{1-x}$In$_x$Te is a $s-$wave superconductor.

\section{Method}

The mirror Chern number has been computed
using the hybrid Wannier charge center scheme \cite{PhysRevB.83.235401,Z2pack},
where the Wannier functions are constructed from first-principles calculations.
The band structure and the projection of the mirror eigenvalues on the 
topological surface states have been computed using
the   density
functional theory and  generalized gradient  approximation for
the         exchange         and        correlation         functional
\cite{PhysRevLett.77.3865}. Fully relativistic pseudopotentials within
the  projector augmented  wave  (PAW) scheme  \cite{PhysRevB.50.17953}
have  been  used self-consistently within plane wave  basis set  with the 
kinetic energy  cut-off of 340~eV.  
We  used  the Vienna  Ab  initio Simulation Package (VASP) for band structure 
and mirror projections \cite{KRESSE199615,PhysRevB.54.11169}, and
{\sc Quantum ESPRESSO} \cite{QE-2009} to compute the Wannier functions.
 Both SnTe and Sn$_{1-x}$In$_x$Te  were modelled
in  the  rock-salt structure.   We  used  the experimental  lattice
parameter  for  SnTe, and  following  the  lattice variation  observed
experimentally   \cite{PRB87-140507,PRB93-024520} making   a   linear
reduction of the lattice parameter  for the In content compounds.  The 
Brillouin  zone  (BZ)  was  sampled   by  using  the
Monkhorst-Pack       special       ${\bm       k}$-points       scheme
\cite{PhysRevB.13.5188}, with grid-sizes as discussed later.

Calculations  for bulk  Sn$_{1-x}$In$_x$Te systems  were performed  by
using a 64-atom  cubic unit cell containing 32 cations  (Sn/In) and
32  anions (Te)  at the  rock-salt  basis sites.   For $x=0,  0.03125,
0.125$ and 0.25 the  number of In dopants per cell was 0,  1, 4 and 8,
respectively.    The     BZ    sampling    was    done     with    the
(10$\times$10$\times$10) mesh of ${\bm k}$-points.

To compute  surface states  we modelled the  Sn$_{1-x}$In$_x$Te system
using two  types of cells containing  an atomic slab and  a minimum of
15~{\AA} of vacuum region.  For $x=0$ (pristine SnTe) and $x=0.125$ we
used a  tetragonal cell  with the  atomic slab  containing 128  and 64
atoms, respectively. For $x=0.125$ each slab in a tetragonal unit cell
contained 4 In, 28 Sn and  32 Te atoms.  For $x=0.03125$ we considered
a cubic  slab with  256 atoms  (4 In,  124 Sn  and 128  Te) in  a unit
cell. The BZ sampling for the cubic and tetragonal cells was done with
(10$\times$10$\times$1) and (7$\times$7$\times$1) ${\bm
  k}$-point meshes, respectively.

\begin{figure}[ht] 
    \centering
    \includegraphics[width=10cm,height=8.3cm]{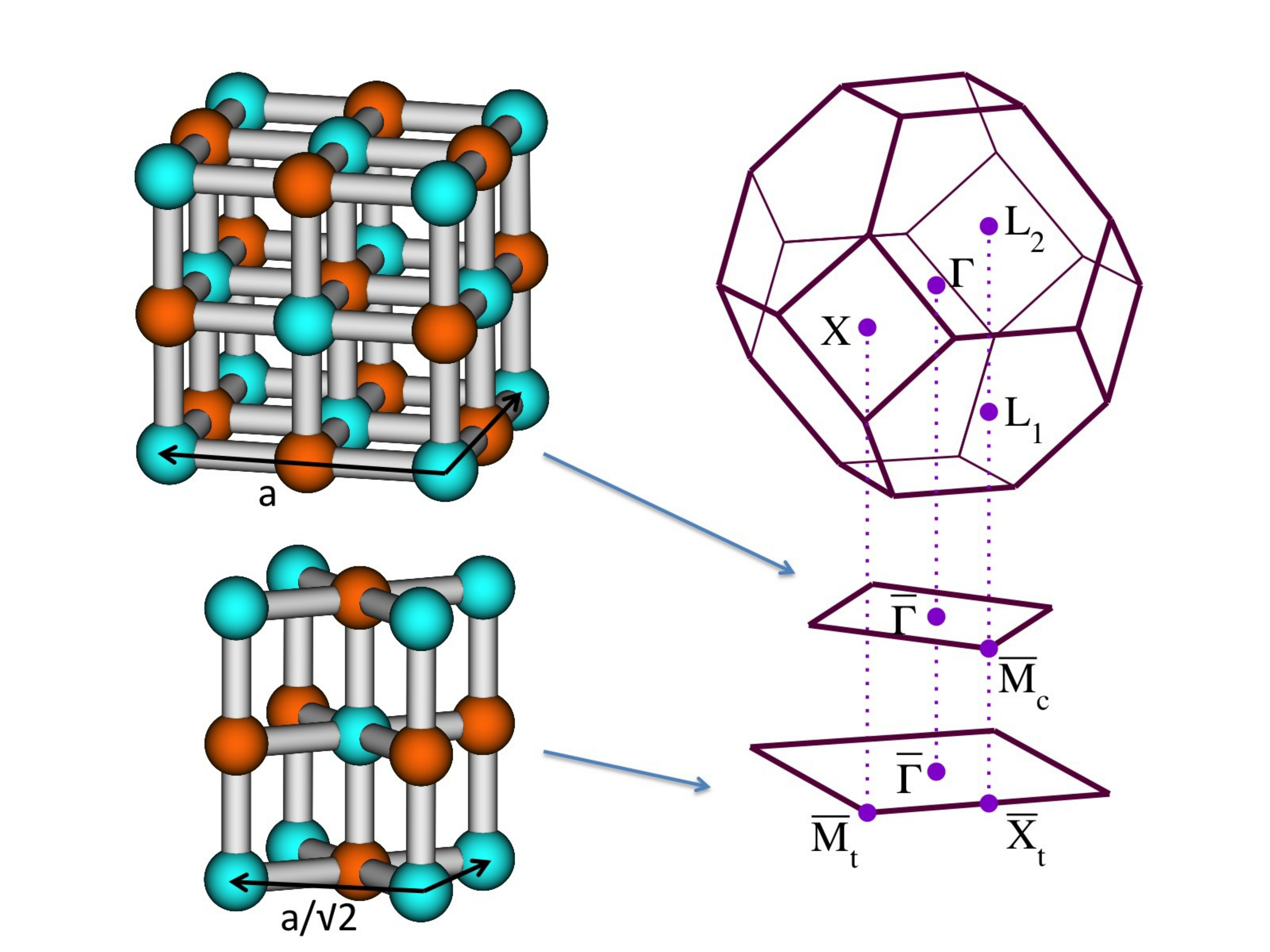}
    \caption{The rock-salt  structures of SnTe modelled  in this work.
      The panel on the left shows  a sample cubic cell of size $a$ and
      a sample tetragonal unit  cell of base $(a/\sqrt 2\times a/\sqrt
      2)$. The  panel on  the right  hand side shows  the BZ  for bulk
      (fcc) on the top, for  the sample cubic structure in the middle,
      and for the sample tetragonal structure at the bottom. Note that
      the middle  and bottom projections  on the right hand  side show
      $(1\times  1)/a$  and   the  $(\sqrt  2\times  \sqrt  2)/a$
      projected reciprocal unit cells.}
    \label{cells}
\end{figure}

In order  to discuss symmetry-related  features in the  band structure
results  for  both bulk  and  slab  geometries,  we have  presented  a
schematic sketch  of a sample cubic cell  of size $a$ and  of a sample
tetragonal cell of base $(a/\sqrt 2\times a/\sqrt 2)$ on the left hand
side of Fig.  \ref{cells}. On  the right hand side of Fig. \ref{cells}
we  have shown the  correspondence between  the important  BZ symmetry
points for  bulk (fcc) and their  projection on the BZs  for the cubic
and  tetragonal structures used  in our  modelling. In  particular, we
note that the  bulk L points map onto the ${\bar  M}_{\rm c}$ point of
the  cubic structure  and onto  the ${\bar  X}_{\rm t}$  point  of the
tetragonal structure.

\section{Results and Discussions}

\subsection{Topological Phase for In Doped Bulk SnTe}

\begin{figure}[ht!] 
    \centering
 \includegraphics[width=\columnwidth,keepaspectratio=true]{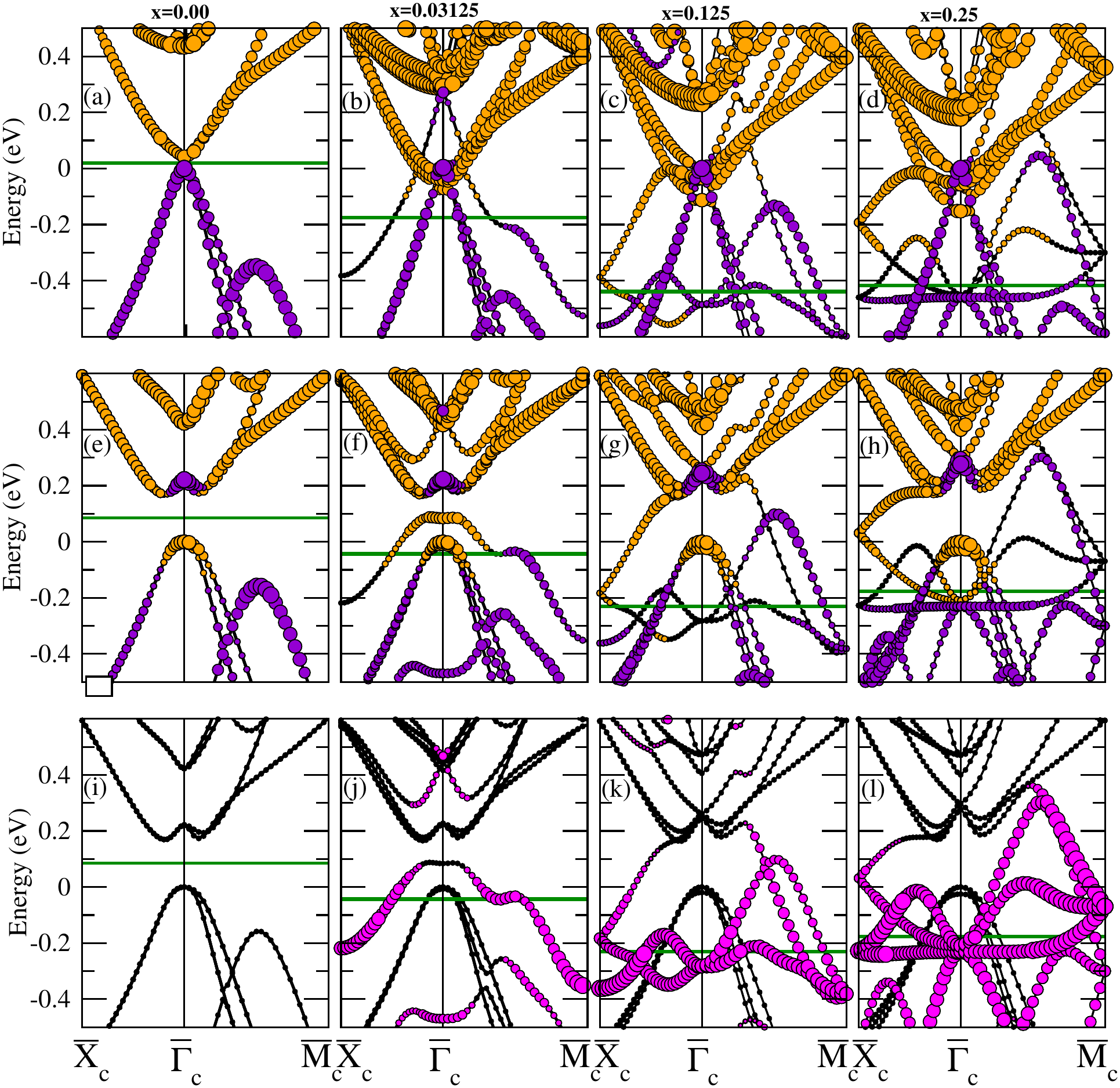}
    \caption{Band  structure without  SO coupling  (top
        row (a)-(d)) and with SO (middle row (e)-(h)) showing the band inversion for
      the projected orbitals  $\Phi_{+} = [\phi^{Te}_{P}+\phi^{Sn}_S]$
      (in  purple)  and  $\Phi_{-}  =  [\phi^{Sn}_P+\phi^{Te}_S]$  (in
      orange). And bottom row (i)-(l) is the projection of In-5s  electronic states
      of  pristine   SnTe, Sn$_{0.96875}$In$_{0.03125}$Te, Sn$_{0.175}$In$_{0.875}$Te,
      and Sn$_{0.75}$In$_{0.25}$Te, respectively.  The symbol sizes are proportional
      to the projection intensities. The horizontal green
      line is the Fermi level. }
    \label{bulks}
\end{figure}

At room  temperature SnTe has  the simple rocksalt  crystal structure.
This  IV-VI compound  is a  narrow-gap (E$_g$=0.18  eV) semiconductor,
with  the  conduction and  valence  band  edges  located at  the  four
equivalent  L  points in  the  fcc  BZ (two  of  which  are marked  in
Fig. \ref{cells}). In  our modelling we chose to  use the experimental
lattice constant of  6.321 {\AA} in order to  facilitate comparison of
electronic states in  the gap region.  In Fig.  \ref{bulks}(a) we have
plotted  the electronic  band structure  of bulk  SnTe using  a simple
cubic (2$\times$2$\times$2) unit cell (8 times the volume of the cubic
cell in Fig \ref{cells}) as mentioned in the Method section.  The L point of the fcc
BZ  is folded  onto the  zone  centre $\bar\Gamma$  for this  periodic
structure.  As  shown in the  bottom panel of Fig.   \ref{bulks}(a) we
obtain a band gap of 0.17 eV  for bulk SnTe when the SO interaction is
included.  

In the asbence of  SO coupling, the VBM and CBM states  at the L point
in the BZ  for IV-VI compounds are contributed  by (s-cation, p-anion)
and (s-anion,  p-cation), respectively \cite{PRB-3-1254-71}.   This is
the  expected normal ordering  of the  band edges  in III-V  and II-VI
compunds  \cite{PRB-4-1877-71,pssb-112-581-82}.   Due to  relativistic
interactions the cation/anion characters are inverted in SnTe, and the
projected      orbital     contributions      are      $\Phi_{-}     =
[\phi^{Sn}_P+\phi^{Te}_S]$  (in  orange)   for  VBM  and  $\Phi_{+}  =
[\phi^{Te}_{P}+\phi^{Sn}_S]$   (in  purple)   for   CBM.   The   Bloch
eigenstates on the $\Gamma~L_1~L_2$ plane  in the fcc BZ are invariant
under the  mirror symmetry of  rocksalt structure with respect  to the
family of  \{110\} planes.   The Bloch  eigenstates for
such  mirror  planes  are  also  eigenstates of  the  mirror  symmetry
operator ${\cal M}$,  with eigenvalues $\pm i$. 
For each class of Bloch eigenstates we compute the mirror
Chern number using the Wannier charge center technique \cite{PhysRevB.56.12847}
obtaining $n_M  = -2$ in agreement with previous results 
\cite{HsiehNC-2012,Tanaka2012}.

For Sn$_{1-x}$In$_x$Te the bulk
conduction  and valence  bands edges  (CBM and  VBM) for  $x=0.03125$,
$x=0.125$ and $x=0.25$  are still inverted at the $\bar\Gamma$(L) point, 
as  we can see in Fig.~\ref{bulks}-(f)-(h). However for low In content
$x=0.03125$ there is a clear total band gap, for higher In concentration
it becomes a gapless system. 
Without the  inclusion of  spin-orbit interaction the  conduction band
dips inside the valence  band, making Sn$_{1-x}$In$_x$Te metallic (see
Fig.~\ref{bulks}-(b)-(d)).  This  is in contrast to  the pristine case
($x=0$), where there is still a very small gap between the VBM and CBM
(Fig.~\ref{bulks}-(a)) without the inclusion of the
spin-orbit interaction.

For low In doping, $x=0.03125$,  we find  an  In state 
inside the bulk  band gap and the Fermi level  lies slightly below the
valence band maximum (VBM)  (see Fig.~\ref{bulks}-(f)). This makes the
system  a  gapped   p-type  doped semiconductor,  in
agreement  with  a  recent  study \cite{PRB93-024520}.   In  order  to
understand the  In effects  we have highlighted
(in  pink) the In-5s  contribution to  the band  structure (identified
within the  range -0.6  eV up to  +0.5 eV)  in Fig.~\ref{bulks}(j).
Some of  these features are close  to the $\bar\Gamma$  point and some
away  from $\bar\Gamma$  but close  to $E_F$.   However, close  to the
$\bar\Gamma$ point there is a negligible In-5s contribution in the SnTe
band gap region.

By increasing the In concentration to 12.5\%, Fig.~\ref{bulks}-(k),
the In-derived levels  go inside the valence band  at the $\bar\Gamma$
point. There is a clear  development of In-5s related bands inside and
across  the band  gap. More  importantly, there  are  more significant
(cf. larger  symbols) In-5s contributions just below  $E_F$ across the
BZ.  Similarly  for the high  $x$ regime of  25\% of In,  the metallic
character is  enhanced with more  In-5s levels connecting  valence and
conduction bands as shown in Fig.~\ref{bulks}-(l). 
It is interesting to note that
there is  a flat band originating  from In-5s orbitals  quite close to
the  Fermi level  (around  $E_f-0.054$ eV).   For  both $x=0.125$  and
$x=0.25$  several In-5s  bands  are empty  around  $E_F$, with  larger
contributions for  $x=0.25$.  This  result can explain  the transition
from hole-like carriers in low In doping to electron-like carriers for
higher   In   concentrations    observed   recently   in   experiments
\cite{PRB98-054503}. This transition can
occur when the In-5s impurity  level is partially occupied, turns it
into In$^{+3}$, instead  of the expected In$^{+1}$ state.   This $p-$ to
$n-$type  transition has  been  found to  occur  even for  a lower  In
concentration \cite{PRB93-024520}, which can  be attributed to the
presence  of other  defects like  Sn  vacancies that  lower the  Fermi
level. We  will  return   to  discuss  this  point  by
examining the electronic density of states later. 

\begin{figure}[ht!] 
    \centering
    \includegraphics[width=\columnwidth,keepaspectratio=true]{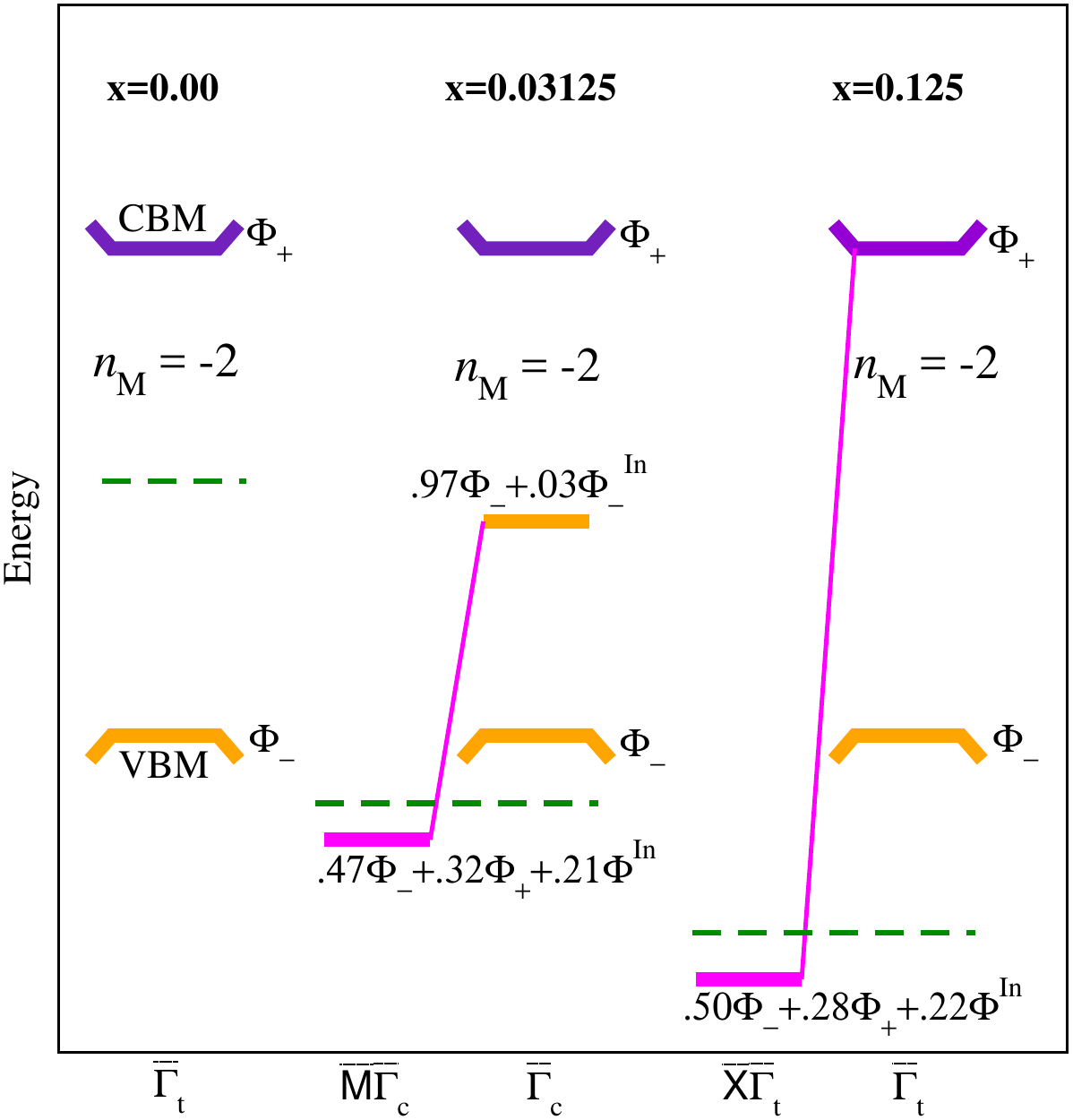}
     \caption{Schematic  illustration  of the  computed
         orbital components of states in and around the  band gap region
         for   Sn$_{1-x}$In$_x$Te,   with   $x=0$,  $x=0.03125$,   and
         $x=0.125$. The  $\Phi_{\pm}$
         states are explained  in the text.  The $\Phi^{\rm  In}$ is a
         mixture of  $\Phi_{\pm}^{\rm In}$.  The Fermi  level is shown
         with green dashed lines.
     }
    \label{sketch}
\end{figure}

The  p-orbitals of cation and  anion has been  used as a
 basis  to  compute  the   mirror  Chern  number  for  pristine  SnTe
 \cite{HsiehNC-2012},     as     well     as     IV-VI     monolayers
  \cite{PRB93-161101}.  The p-orbitals  of cations contributing to VBM
  and the  p-orbitals of anions  contributing to CBM, are  all aligned
  along the [111] direction.   This inverted orbital ordering produces
  a negative band gap.  In Fig.~\ref{sketch} we show schematically the
  evolution of  the VBM  and CBM orbital  characters when In  ions are
  incorporated at  Sn sites.   We can still  identify the  $\Phi_{-} =
  [\phi^{Sn}_P+\phi^{Te}_S]$          and          $\Phi_{+}         =
  [\phi^{Te}_{P}+\phi^{Sn}_S]$ for the  VBM and CBM, respectively, but
  we have an  additional In state which is always  around the band gap
  region.  The  In contribution of  this state is  quite low at  the L
  (${\bar\Gamma}_t$)  point  but   increases  away  from  this  point.
  Importantly,  the In  state keeps  the  same character  as the  VBM,
  wheather it is inside the band gap or above the CBM.
For low In concentration, $x=0.03125$,  the In state is located inside
the band gap and its  contribution is pure p-orbital ($\Phi_-^{In}$) aligned along the
[111] direction, the same character as that for the VBM cation Sn ion.
As the gap is still  inverted, the non-trivial topology is maintained.
For high In concentration the In state is resonant inside the conduction band, and
both VBM and CBM characters are the same as for the pristine one. 
Using the same procedure as described above the computed mirror Chern number 
for each of $x=0.03125$ and 0.25 is $n_M=-2$.

\subsection{Topological Surface States}

In  the  previous  sub-section  we  have  clearly  verified  the  band
inversion and non-trivial topological characteristics of pristine SnTe and
Sn$_{1-x}$In$_x$Te. 
In  a recent experimental study,  using ARPES measurements,
Sato  {\it  et  al}  \cite{PRL110-206804} reported  the  existence  of
topological surface states, leading  to the possibility of topological
superconductivity in  Sn$_{1-x}$In$_x$Te.   
In order to clarify the topological nature,  we also performed  calculations with  a
repeated  slab  geometry for  pristine  ($x=0$)  and  In doped  system
containing  a low ($x=0.03125$)  and a  higher In  doped concentration
($x=0.125$).  Details  of the slab makeup, including  number of atomic
layers  and vacuum  region, have  been presented  earlier. 
The mirror symmetry has been maintained for each slab geometry.
However, substitutional In impurity  breaks the top and bottom surface
symmetries,  yielding a  reduction  in the  hybridization between  the
topological states of opposite surfaces.

\begin{figure}[ht!] 
    \centering
    \includegraphics[width=\columnwidth,keepaspectratio=true]{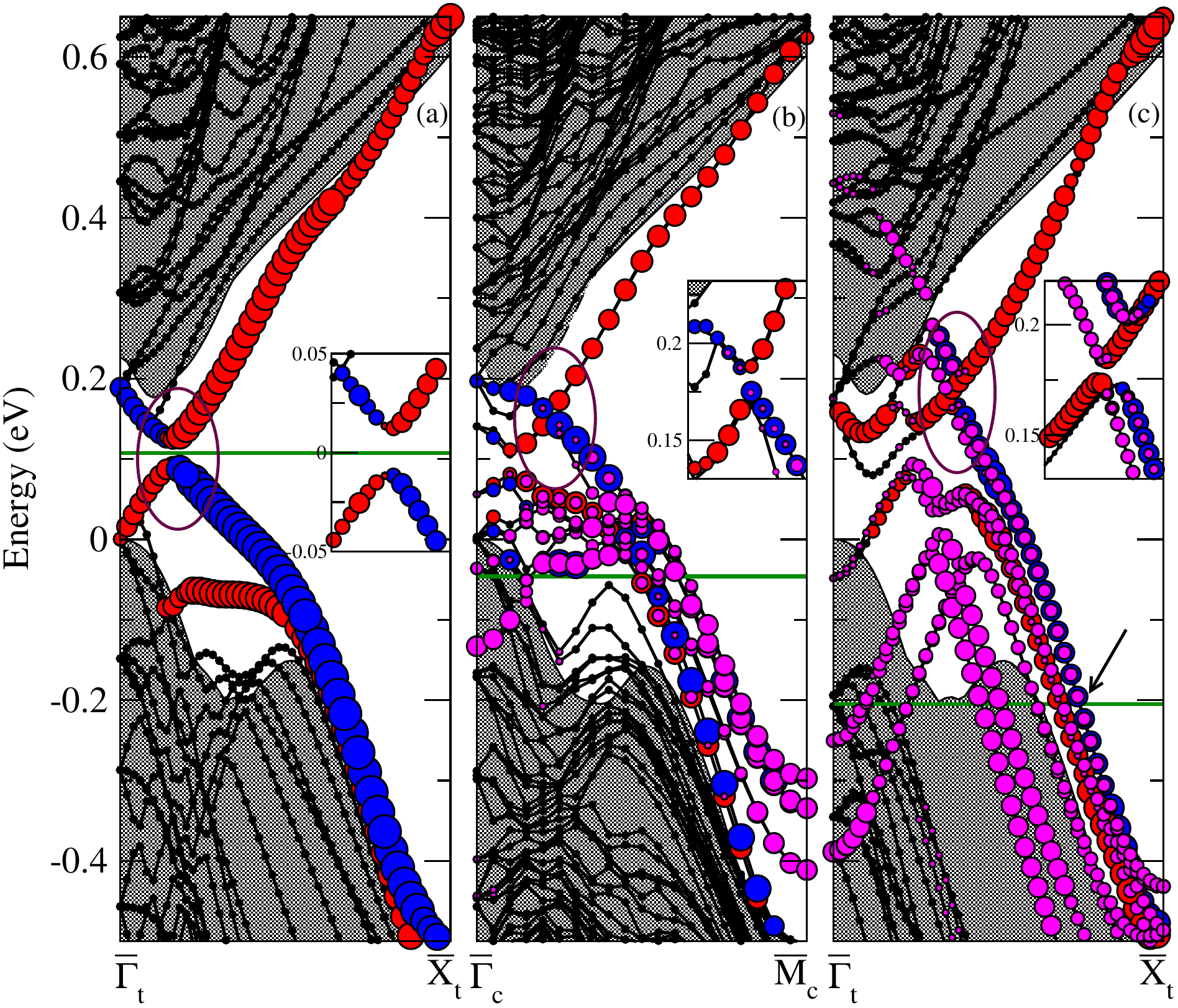}
     \caption{Topological surface  states for  Sn$_{1-x}$In$_x$Te: (a)
       $x=0$ (pristine  SnTe); (b)  $x=0.03125$; (c)  $x=0.125$.  Blue
       and red symbols represent the  mirror eigenvalues $+i$ and $-i$
       projected in half  of the slab cell for bands around the valence and
       conduction band edges. The pink are the In-5s projected bands. The          
       horizontal
       green line  is the Fermi level.  The back shadow region  is the
       bulk projected  band. The insets show magnified  versions around the
       Dirac crossing region.}
    \label{ss}
\end{figure}

Expressing,  with   respect  to   the  mirror  plane,   symmetric  and
antisymmetric real  space parts  of wavefunction  as $|s\rangle$  and $|a\rangle$,
respectively,  and the  spinor  as $|\sigma\rangle$,  the mirror  eigenvalue
equations can be stated as
$${\cal  M}|s,\pm\sigma\rangle  =  \pm i  |s,\pm\sigma\rangle;  \qquad
{\cal M}|a,\pm\sigma\rangle =\mp i |a,\pm \sigma\rangle.$$ This allows
for two  sets of  topological surface states  to be  realized: the state
$(|s,\sigma\rangle+|a,-\sigma\rangle)$ corresponding to the eigenvalue
$+i$ and the state $(|s,-\sigma\rangle+|a,\sigma\rangle)$ corresponding
to the  eigenvalue $-i$.   Panel (a) in  Fig.~\ref{ss} shows  the band
structure of  the pristine ($x=0$)  SnTe slab geometry. The  bulk band
structure projected  on to the tetragonal  slab BZ is shown  as shaded
region.  Consistent  with the  mirror symmetry  and band  inversion, a
pair of surface bands corresponding  to the mirror eigenvalues $\pm i$
inside the bulk band gap region  are identified (and shown in blue and
red   colors)  extending   along  the   ${\bar\Gamma}_t$-${\bar  X}_t$
direction of  the tetragonal BZ.   The tiny  band gap (around  36 meV)
between the  two surface  bands in  the middle of  the bulk  band gap,
instead of Dirac  crossing, is due to the interaction  between the top 
and bottom surfaces of the slab. We also  clearly identify  a surface
state (non  topological origin) lying  at around  $E_F - 0.17$  eV for
part of the ${\bar\Gamma}_t$-${\bar X}_t$ direction.

As we  can see  in panels  (b) and  (c) of  Fig.~\ref{ss}, topological
surface states  are present  for both the  low ($x=0.03125$)  and high
($x=0.125$)  In   concentration  regimes.   The  mirror   symmetry  is
preserved for  both In concentrations, as required for the presence
of symmetry protected surface states.   It is interesting to note that
the topological state branch (blue color) coming from the valence band
has contributions from In orbitals (pink color), while the topological
state  coming from  the  conduction  band (red  color)  has pure  SnTe
character, similar to the pristine case in panel (a).

In a  previous work in TCI  \cite{SciRep8-9452} it was  found that, as
the slab  thickness increases the  interaction between top  and bottom
topological  surface states  reduces, leading  to a  reduction  in the
surface states band gap.  We find that the surface state gaps for both
$x=0.03125$ and $x=0.125$ are smaller  than that for $x=0$.  Since the
pristine  SnTe  slab, and  the  slab  for  $x=0.03125$ have  the  same
thickness,  we  conclude that  the  presence  of  In impurity  further
reduces  the surface  band  gap.  The  presence  of substitutional  In
breaks the top and bottom  surface symmetries, yielding a reduction of
the hybridization between the topological states of opposite surfaces.
Our work  reveals that  Indium doping of  SnTe shifts the  Dirac crossing
energy  position  and  also   affects  the  Fermi  velocities  of  the
topological surface  states.  The  main changes occur  for topological
hole states.  As the doping moves the Fermi level downwards, the Fermi
velocity increases with respect to the pristine system.


In Table I  we summarise the topological surface  state parameters for
Sn$_{1-x}$In$_x$Te.  For  pristine SnTe,  the Dirac crossing  point is
identified at a location with wavenumber $k_D(L)$ away from the bulk L
point. For  Sn$_{1-x}$In$_x$Te the Dirac crossing  point moves further
way from  the L  point as $x$  increases.  The Fermi  wavenumber $k_F$
also increases as $x$  increases. Our computed values of  $k_F$ are similar
to     the      values     reported     in      experimental     works
\cite{PRB98-054503,PRB93-075132},  {\it   albeit}  for  different  $x$
values and different surfaces.   The electron and hole velocities have
been computed in a range of  0.2~eV above and below the Dirac crossing
point. The electron velocity, $v_e$,  is almost independent of the $x$
values.   It is  found that  the hole  velocity, $v_h$,  has increased
significantly for $x=0.125$.  In  contrast, the Fermi velocity of the
topological states, $v_F$,
increases  steadily   with  increase  in  $x$.    Our  computed  Fermi
velocities  are  in  reasonable  agreement with  those  obtained  from
quantum        oscillation        and        ARPES        measurements
\cite{PRB98-054503,PRB93-075132}. In particular, the increase in $v_F$
with  increase  in   $x$  has  also  been  revealed   from  the  ARPES
measurements in \cite{PRB93-075132} for  $x=0.23$ and $x=0.41$ for the
Sn$_{1-x}$In$_x$Te(111) surface.

\begin{table}[htb]  
\centering
\caption{Topological  surface state  parameters:  k$_D(L)$ is the  Dirac
  crossing shift with respect to the pristine L point, Fermi wavenumber k$_F$,
  electron  velocity v$_e$,  hole velocity  v$_h$, and  Fermi velocity
  v$_F$. Wavenumbers in units  of {\AA}$^{-1}$, and the velocities are
  in units of m/s.}
\begin{tabular}{|c|c|c|c|c|}
\hline \hline
& Pristine & $x=0.03125$ & $x=0.125$ & Experimental reports\\
&              &                      &                   & ($x$ in the range 0.23-0.41)\\
\hline
k$_{D}$(L) & 0.056 & 0.077 & 0.121 & \\
k$_F$ & 0.000 & 0.101 & 0.131 & 0.04{\cite{PRB98-054503}}, 
0.06-0.10{\cite{PRB93-075132}}  \\
v$_e$ & 3.02 & 3.03 & 2.76 & \\
v$_h$ & 1.85 & 1.82 & 3.86 & \\
v$_F$ & 2.46 & 3.56 & 5.79 &  2.5{\cite{PRB98-054503}}, 
5.8-6.0{\cite{PRB93-075132}} \\
\hline \hline
\end{tabular}
\label{table1}
\end{table}

We have estimated the Dirac crossing  point to lie at 220~meV and 
380~meV
above   the  Fermi   level   for   $x=0.03125$  and   $x=0.125$,
respectively. The relative increase  in $E_D-E_F$ with increase in $x$
has  also been  reported  in \cite{PRB93-075132}  from a  quantitative
analysis of the ARPES spectrum for the (111) surface with $x=0.23$ and
$x=0.41$. Bulk states are always inside the topological surface bands,
{\it  i.e.} with  k(bulk)$<$k$_D$(surf). This  is consistent  with the
analysis of the ARPES  study in \cite{PRB93-075132} for their $x=0.23$
and $x=0.41$ samples.

\subsection{Superconductivity Character}

\begin{figure}[ht!] 
    \centering
    \includegraphics[width=\columnwidth,keepaspectratio=true]{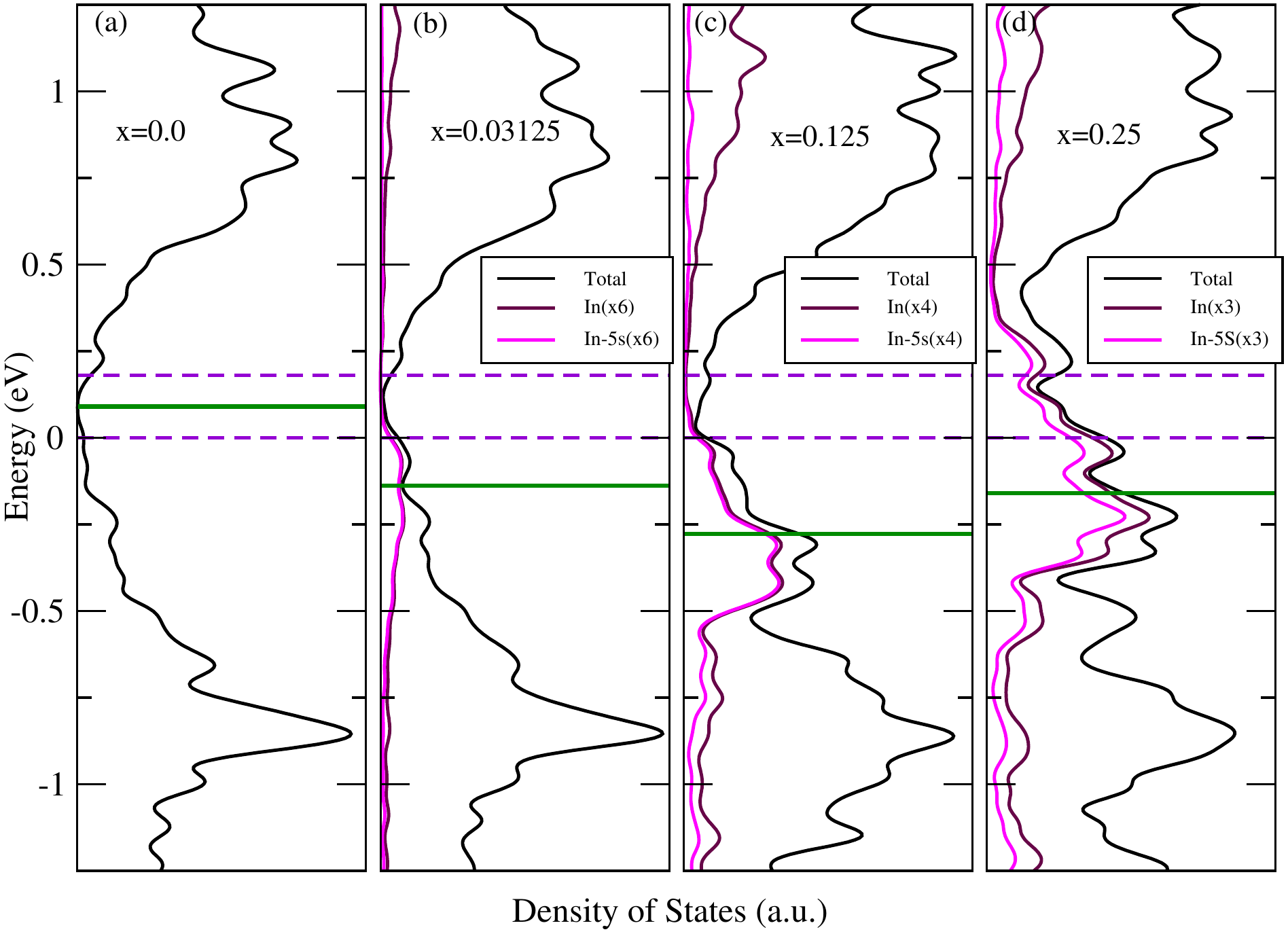}
    \caption{Total  density of  states  (black) and  local density  of
      states projected on In (maroon) and on In-5s orbital (pink). The
      horizontal  solid (green)  line  is the  Fermi  energy, and  the
      dashed  (violet)  lines  are  the valence  and  conduction  band
      edges. }
    \label{dos}
\end{figure}

Transport, magnetization, and heat  capacity measurements show that In
substitution  at Sn  sites turns  the TCI  SnTe into  a superconductor
\cite{PRB87-140507}.    The   superconductivity   character  of   bulk
Sn$_x$In$_{1-x}$Te can  also be qualitatively analysed  from our first
principles   calculations.    The   electronic   density   of   states
distribution near  the Fermi level, $DOS(E_F)$, plays  a dominant role
in forming the  BCS type superconducting state.  According  to the BCS
theory, the formation energy of  a Cooper pair increases linearly with
$DOS(E_F)$ \cite{Ziman-1964}.  With this in mind, in Fig.~\ref{dos} we
project   the  total   and   In  contribution   to   the  density   of
states. Comparing the results for $x=0$ (pristine SnTe) and the low In
concentration of $x=0.03125$ we find that there is a clear development
of $DOS(E_F)$  for the latter  case.  There is a  progressively larger
development  of $DOS(E_F)$  for  the larger  values  of $x=0.125$  and
$x=0.25$.  Our computed  density of states per unit  cell at the Fermi
level  for the  three doping  concentrations are  0.19, 0.73  and 0.78
states/eV for $x=0.03125$, $x=0.125$ and $x=0.25$, respectively. These
values  are in  quite good  agreement  with the  estimated density  of
states  based on magnetic  susceptibility data  \cite{PRB93-024520} of
0.44 and 1.17 states/eV  for $x=0.05$ and $x=0.40$, respectively.  For
each  of the three  $x$ values  In-5s orbitals  make almost  the total
contribution towards  $DOS(E_F)$.  Our  results thus support  a s-wave
superconductivity  in  Sn$_x$In$_{1-x}$Te,  in agreement  with  recent
experimental                                               observations
\cite{EPL108-37010,PRB90-064508,NL15-3827,PRB96-104502,PRB97-024511}.
We  observe also from  Fig.~\ref{dos} that  the In-5s  orbital becomes
less occupied for  higher In doping regime, which  prompts the $p-$ to
$n-$type  transition as  a  function of  increasing In  concentration,
providing  support to  the  analysis of  experimental measurements  in
\cite{PRB93-024520,PRB98-054503}.

\section{Conclusion}

We report on the non-trivial
topological  properties  of the superconductor Sn$_{1-x}$In$_x$Te.   
The  band  inversion
characteristic, {\it  viz.} swapping of the  normal contributions from
p-anion  and s-cation  orbitals for  the valence  and  conduction band
edges (VBM and  CBM), has been identified to arise  in the presence of
spin-orbit  interaction.  This  inversion  is preserved  for In content
systems with $x=0.03125, 0.125$ and 0.25. The non-trivial phase has been
confirmed by computing the mirror Chern number which is non-null for all
those In concentrations.
Topological states with opposite $\pm i$ mirror 
eigenvalues has been identified on the (001) surface of the Sn$_{1-x}$In$_x$Te.
With increase in In concentration, the location of the surface
Dirac crossing  point shifts progressively  towards larger wavenumber.
Also,  the  doping moves  the  Fermi  level  downwards and  the  Fermi
velocity progressively and significantly increases as $x$ increases. 
We also verified that the density of states at
the Fermi  level for finite $x$  values has been found  to come mostly
from  In 5s orbitals. This  confirms  the  s-wave  nature  of  the
superconducting state in agreement with some experimental studies.
The results  presented in this work demonstrate a non-trivial topological
phase for the superconductor Sn$_{1-x}$In$_x$Te, and provide a detailed
description of the topological state properties.

\section*{Acknowledgements}

One of  the authors (TMS)  acknowledge the financial support  from the
Brazilian Agencies INCT in Carbon Nanomaterials, CNPq, CAPES, FAPEMIG,   and the computational facilities of LNCC and Cenapad.

\bibliography{snte}
\bibliographystyle{unsrt.bst}
\end{document}